\documentstyle[aps,preprint]{revtex}

\newcommand{\tr}{\mbox{Tr} }
\newcommand{\ket}[1]{\left | #1 \right \rangle}
\newcommand{\bra}[1]{\left \langle #1 \right |}

\newcommand{\proj}[1]{\ket{#1} \! \bra{#1}}

\newcommand{\superop}{\mbox{$\cal E$}}

\begin{document}

\tightenlines

\title{Quantum privacy and quantum coherence}
\author{Benjamin Schumacher}
\address{Department of Physics, Kenyon College, Gambier, OH  43022  USA}
\author{Michael D. Westmoreland}
\address{Department of Mathematical Sciences, Denison University,
         Granville, OH  43023  USA}

\maketitle

\begin{abstract}
We derive a simple relation between a quantum
channel's capacity to convey coherent (quantum) information and 
its usefulness for quantum cryptography.
\end{abstract}


A quantum communication channel can be used to perform a variety
of tasks, including:
\begin{itemize}
	\item  Conveying classical information from a sender to
		a receiver.
	\item  Conveying quantum information (including quantum
		entanglement) from a sender to a receiver.
	\item  Creating shared information between a 
		sender and receiver, information that is
		reliably secret from any third party and can
		thus be used as a cryptographic key for later 
		private communication.  (The use of quantum
		channels to aid in cryptographic tasks such as
		key distribution is called quantum cryptography.)
\end{itemize}
Each of these tasks can be performed in the presence of noise.
Indeed, in quantum cryptography the noise is of central importance
in revealing the activity of an eavesdropper.

Deutsch {\em et al.} \cite{qprivacy} examined the security of quantum
cryptographic schemes over quantum channels that contain noise.
They pointed out that any protocol which allowed ``entanglement
purification'' between two parties automatically provided a means
of communicating secret information that no third party could share.
Here we will continue this line of thought by showing that the
privacy of the channel, measured by the amount of information
available to the receiver that is not available to any eavesdropper,
can be made at least as great as the channel's {\em coherent 
information} \cite{coherent}.

Suppose Alice prepares a quantum system $Q$ in an initial state
$\rho^{Q}$.  
Alice conveys the system $Q$ through a noisy quantum channel
to Bob.  The noisy channel may be described by a superoperator
$\superop^{Q}$, so that the final state $\rho^{Q'} = \superop^{Q}
(\rho^{Q})$.

The evolution of the channel given by the superoperator $\superop^{Q}$ 
is in fact unitary evolution on a larger quantum system that includes
the environment $E$ of the system.  This environment may be considered
to be initially in a pure state $\ket{0^{E}}$.  In this case, the
superoperator is given by
\begin{equation}
	\superop^{Q} (\rho^{Q}) = \tr_{E} U^{QE} \left (
				  \rho^{Q} \otimes \proj{0^{E}} \right )
				  {U^{QE}}^{\dagger} .
\end{equation}
We can assume that the environment is initially in a pure state without
any loss of generality, since we can always imagine that a ``local''
environment in a mixed state is just part of a larger system in a pure
entangled state.

We imagine first that the initial mixed
state $\rho^{Q}$ of $Q$ arises from $Q$'s entanglement
with some other ``reference'' system $R$ in Alice's possession.
Alice's goal in sending $Q$ to Bob is to establish some
quantum entanglement between her reference system $R$ and
Bob's output system $Q'$.  That is, Alice is sending {\em
quantum information} via the channel to Bob.

As discussed in \cite{entex},
the entropy exchange $S_{e}$ measures the amount of information that 
is exchanged between the system $Q$ and the environment $E$ during
their interaction.  If the environment is initially in a pure state,
the entropy exchange is just the environment's entropy after the
interaction---i.e., $S_{e} = S(\rho^{E'})$, where $\rho^{E'}$ is the
final state of $E$.  (The entropy here is just the ordinary von Neumann
entropy of a density operator, $S(\rho) = - \tr \rho \log \rho$.)
The entropy exchange is entirely determined by the initial state
$\rho^{Q}$ of $Q$ and the channel dynamics superoperator $\superop^{Q}$;
that is, the entropy exchange is a property ``intrinsic'' to $Q$ and
its dynamics.

The coherent information $I_{e}$, introduced in \cite{coherent}, 
is given by
\begin{equation}
	I_{e} = S(\rho^{Q'}) - S_{e} .
\end{equation}
The coherent information has many properties that suggest it as the 
proper measure of the quantum information conveyed from Alice to Bob
by the channel.  For example, $I_{e}$ can never be increased by 
quantum data processing performed by Bob on
the channel output, and perfect quantum error correction of the
channel output is possible for Bob if and only if no coherent 
information is lost in the channel \cite{coherent}.  
The coherent information seems to
be related to the capacity of a quantum channel to convey quantum
states with high fidelity \cite{qcapacity}.

Alice might on the other hand be using the channel to send classical 
information to Bob.  Alice prepares $Q$ in one of a set of possible
``signal states'' $\rho^{Q}_{k}$, which are used by Alice with
{\em a priori} probabilities $p_{k}$.  The average state $\rho^{Q}$
is given by
\begin{equation}
	\rho^{Q} = \sum_{k} p_{k} \rho^{Q}_{k} .
\end{equation}
Bob receives the $k$th signal as $\rho^{Q'}_{k} = 
\superop^{Q}(\rho^{Q}_{k})$.  Because the superoperator is linear,
the average received state is
\begin{equation}
	\rho^{Q'} = \sum_{k} p_{k} \superop^{Q}(\rho^{Q}_{k})
                  = \superop^{Q} (\rho^{Q}).
\end{equation}
Bob attempts to decode Alice's message (that is, to identify
which signal state was chosen by Alice) by measuring some
{\em decoding observable} on his received system $Q'$.

The amount of classical information conveyed from Alice to Bob,
which we will denote $H_{Bob}$,
is governed by the quantity $\chi^{Q'}$, defined by
\begin{equation}
	\chi^{Q'} = S(\rho^{Q'}) - \sum_{k} p_{k} S(\rho^{Q'}_{k}) .
\end{equation}
This quantity is significant in two ways:
\begin{itemize}
	\item  $H_{Bob} \leq \chi^{Q'}$,
		regardless of the decoding observable chosen
		\cite{holevo,ssubadd}.
	\item  $H_{Bob}$ can be made as close as desired to
		to $\chi^{Q'}$ by a suitable choice of code
		and decoding observable.  To make $H_{Bob}$ near
		$\chi^{Q'}$, Alice must
		in general use the channel many times and employ
		code words composed of many signals; Bob must
		perform his decoding measurement on entire code
		words.  The net result is that the channel is
		used $N$ times to send up to $N \chi^{Q'}$ bits
		of classical information reliably \cite{ccapacity}.
\end{itemize}
In short, $\chi^{Q'}$ represents an upper bound on the 
classical information conveyed from Alice to Bob, an upper bound
that may be approached arbitrarily closely if Alice and Bob
use the channel efficiently.

If this general picture is used to describe a quantum cryptographic 
channel, then the eavesdropper (``Eve'') must be supposed to have
access to some or all of the environment system $E$ with which
$Q$ interacts.  In other words, 
the environment includes any apparatus used by Eve to
gather information about Alice and Bob's communication.  The
evolution superoperator $\superop^{Q}$ thus describes all of the
effects of the eavesdropper on the channel; or, to put it 
another way, all of the eavesdropper's efforts at ``tapping''
the link between Alice and Bob are contained in the interaction
operator $U^{QE}$.  The information $H_{Eve}$
available to the eavesdropper
will be limited by 
\begin{equation}
	\chi^{E'} = S(\rho^{E'}) - \sum_{k} p_{k} S(\rho^{E'}_{k}).
\end{equation}
The limitation $H_{Eve} \leq \chi^{E'}$ holds whether or not Eve has
access to the entire environment.  If Eve can only see a subsystem
$D$ of the full environment, then we can make the stronger statement
$H_{Eve} \leq \chi^{D'}$, where $\chi^{D'} \leq \chi^{E'}$ 
\cite{ssubadd}.

We define the ``privacy'' $P$ of a channel to be 
\begin{equation}
P = H_{Bob} - H_{Eve}.  
\end{equation}
This definition makes sense, because any positive difference
$H_{Bob} - H_{Eve}$ can be exploited by Alice and Bob 
to create a reliably secret string of key bits of length 
about $P$ \cite{secretkey}.

Alice and Bob wish to make $P$ as large as possible.
However, they cannot control the actions of the eavesdropper.
Thus, they must assume that the eavesdropper is acquiring
her greatest possible information from the channel.  The
``guaranteed privacy'' $P_{G} = \inf P$, where the infimum
is taken over all of Eve's possible strategies that are
consistent with the superoperator $\superop^{Q}$ describing
the channel.  Since $H_{Eve} \leq \chi^{E'}$, we have
\begin{equation}
P_{G} \geq H_{Bob} - \chi^{E'} .
\end{equation}

Alice and Bob will want to use the channel to make the guaranteed
privacy $P_{G}$ as great as possible.  Let ${\cal P} = \sup P_{G}$ 
be the optimal guaranteed privacy, where the supremum is taken
over all strategies that Alice and Bob may employ to use the
channel.  How big is ${\cal P}$?  As discussed above, by suitable
choice of code and decoding observable, $H_{Bob}$ can be made
arbitrarily close to $\chi^{Q'}$.  Thus,
\begin{equation}
	{\cal P} \geq \chi^{Q'} - \chi^{E'} .
\end{equation}

If Alice and Bob were simply trying to optimize $H_{Bob}$ over
a given noisy channel, it is known \cite{ccapacity} that 
they can do no better than to choose pure states of $Q$ as
the inputs signal states of the channel.  Here, they
are instead trying to maximize the guaranteed privacy $P_{G}$,
so that pure state inputs may not be optimal.  However, we can
certainly find a lower bound for ${\cal P}$ by considering
$\chi^{Q'} - \chi^{E'}$ for pure state inputs.

Assume that the states of $Q$ initially prepared by Alice
are pure states $\ket{\phi^{Q}_{k}}$; also recall that the 
environment $E$ can be presumed to begin in a pure state $\ket{0^{E}}$.
After $Q$ and $E$ interact unitarily, the joint state
$\ket{\Psi^{QE'}_{k}} = U^{QE} \ket{\phi^{Q}_{k}} \otimes \ket{0^{E}}$
will also be a pure state, generally an entangled one.  
The subsystem states, described by density operators
\begin{eqnarray}
	\rho^{Q'}_{k}	& = &	\tr_{E} \proj{\Psi^{QE'}_{k}} \nonumber \\
	\rho^{E'}_{k}	& = &	\tr_{Q} \proj{\Psi^{QE'}_{k}} ,
\end{eqnarray}
will have exactly the same non-zero eigenvalues, so that 
$S(\rho^{Q'}_{k}) = S(\rho^{E'}_{k})$.  Therefore
\begin{eqnarray}
	I^{Q}	& = &	S(\rho^{Q'}) - S_{e}	\nonumber \\
		& = &	S(\rho^{Q'}) - S(\rho^{E'}) \nonumber \\
		& = &	S(\rho^{Q'}) - \sum_{k} p_{k} S(\rho^{Q'}_{k})
			- S(\rho^{E'}) + \sum_{k} p_{k} S(\rho^{E'}_{k}) 
			\nonumber \\
	I^{Q}	& = &	\chi^{Q'} - \chi^{E'} .
\end{eqnarray}
We conclude that
\begin{equation}
	{\cal P} \geq I^{Q} .
\end{equation}
In other words, the ability of the quantum channel to send {\em private} 
information is at least as great as its ability to send {\em coherent}
information.
This result may be viewed as a quantum information theoretic basis for 
quantum cryptography.

It is interesting to note that, although both $\chi^{Q'}$ and $\chi^{E'}$
depend on the choice of pure state inputs for the channel $Q$, the 
difference $\chi^{Q'} - \chi^{E'}$ depends only on the overall 
density operator $\rho^{Q}$ for the inputs.

We have assumed that the properties of the channel, given by the
superoperator $\superop^{Q}$, are known to Alice and Bob.  If 
$\superop^{Q}$ is known, and if $I^{Q} > 0$ for some $\rho^{Q}$,
then the channel may be used to send private information securely.
However, this does not address the question of how Alice and Bob
can establish the necessary properties of the channel without
being deceived by Eve.  

We would like to thank W. K. Wootters and M. A. Nielsen
for helpful conversations and suggestions.

\end{document}